\documentclass[12pt]{article}

\usepackage{epsfig}
\newcommand{\be}{\begin{equation}}
\newcommand{\ee}{\end{equation}}
\newcommand{\bea}{\begin{eqnarray}}
\newcommand{\eea}{\end{eqnarray}}
\newcommand{\beaa}{\begin{eqnarray*}}
\newcommand{\eeaa}{\end{eqnarray*}}
\newcommand{\ben}{\begin{enumerate}}
\newcommand{\een}{\end{enumerate}}
\newcommand{\bi}{\begin{itemize}}
\newcommand{\ei}{\end{itemize}}

\def\zzz{\mathop{\rm Z\kern -0.25em Z}\nolimits}
\def\prt{\partial}
\def\Ba{\phi}

\def\al{\alpha}
\def\ka{\kappa}
\def\prt{\partial}

\def\dl{\delta}
\def\Om{\Omega}

\def\ra{\rightarrow}
\def\half{\frac{1}{2}}

\def\squarebox#1{\hbox to #1{\hfill\vbox to #1{\vfill}}}

\begin{document}

\begin{center}
Phase transition curves for mesoscopic superconducting samples \\[.3cm]
H.T. Jadallah, J. Rubinstein and P. Sternberg\\[.3cm]

\vspace{.2cm}

\begin{minipage}[t]{142.5mm}
\textbf{Abstract. } \textsl {We compute the phase transition
curves for mesoscopic superconductors. Special emphasis is given
to the limiting shape of the curve when the magnetic flux is large.
We derive an asymptotic formula for the ground state of the Schr\"odinger
equation in the presence of large applied flux. The expansion is
shown to be sensitive to the smoothness of the domain. The
theoretical results are compared to recent experiments.}
\end{minipage}

\end{center}
\vspace{.5cm}

\vspace{.5cm}

In their classical experiment, Little and Parks \cite{lipa} considered
the transition between normal and superconducting phases in cylindrical
shells. They discovered that the phase boundary in the $T-H$ plane, where
$T$ is the temperature and $H$ is the applied magnetic field,
is oscillatory.
When the thickness of the shell is very small relative to the
coherence length, the period is approximately the fundamental flux
quantum $\phi_o=h/e^*c$, where $e^*=2e$ is the charge of the Cooper
pair. Taking into account the effect of the finite thickness of the ring, it
can be shown theoretically \cite{grpa} and verified experimentally
(e.g. \cite{lipa},\cite{mos})
that the oscillations in the $T-H$ line
are superimposed on a parabolic background.
The situation becomes more complex if the circular symmetry of
the shell is broken \cite{beru}, but the $T-H$ line is still oscillatory.

The classical literature on the subject assumes that the
superconducting sample is multiply connected, typically with a
ring-like geometry. Nevertheless, an
oscillatory $T-H$ line is also obtained for simply connected
domains. This has been shown experimentally by Buisson et al.
\cite{bui} who considered a small disc. Later Moshchalkov et al.
\cite{mos} and Bruyndoncx et al. \cite{bru2} measured the
transition temperature $T$ as a function of the applied flux for a
mesoscopic square. Both groups reported an oscillatory phase
boundary superimposed on a linear background. The heuristic
reasoning for the oscillations is that as the flux increases, the
wave function concentrates near the boundary, and the sample
appears effectively as a thin ring. This gives rise to a
topological quantization constraint on the phase of the wave
function and hence the Little Parks oscillations. The analogy with
the experiments in thin shells is not perfect, though. In the case
of a simply connected sample under strong fields, the radius of
the 'effective' ring is not fixed. Rather it shrinks as the
applied field increases.

We shall show here that the base $T-H$ curve is parabolic for
small fluxes, linear for large fluxes, and that its precise shape
depends on the smoothness of the domain. We compute the $T-H$
curve for arbitrary mesoscopic two dimensional samples $\Om$. A
constant magnetic field of magnitude $H$ is applied in the
direction orthogonal to the plane. We pay particular attention to
the asymptotic limit in which the flux is large. Near the phase
transition, the induced magnetic field is negligible and
 the wave function is well-approximated by the solution to
 the linearized
Ginzburg Landau equation
\be
\frac{1}{2m}(i h \nabla +\frac{e^*}{c}A)^2\Psi=\al \Psi,
\label{gl}
\ee
(where $\nabla$ is the two-dimensional gradient operator)
with nonreflecting boundary conditions at $\prt \Om$ the boundary of
$\Om$. Here $A$ is the magnetic vector potential satisfying $\nabla \times A
=H\hat{z}$, $m$ is the effective electron mass,
$\al=\frac{\hbar^2}{2m \xi^2_0}\frac{T-T_c}{T_c}$, where $\xi_0$ is the
coherence length at $T=0$ and $T_c$ is the critical temperature
in the absence of magnetic field.
Since we are interested in finding the relation between $T$ and $H$, the
problem is mathematically equivalent to computing the
ground state
of the Schr\"odinger equation for a particle under a
 magnetic field.
The strong field asymptotics can also be shown to be equivalent
to evaluating the critical
magnetic field $H_{c3}$ for materials with large Ginzburg
Landau parameter \cite{bpq}, \cite{best}.

We denote by $\phi$ the applied flux measured in units of
$\phi_0$. When $\phi \gg 1$ we expect $|\Psi|$ to be very small
except at a boundary layer whose nondimensional thickness is
$O(\phi^{-\half})$. The detailed structure of $\Psi$ and the
asymptotic expansion of $\al$ depend upon the geometry and
smoothness of $\Om$.

It is convenient to scale the domain $\Om$ canonically through
a similarity transformation that preserves its shape while
changing its area to $2 \pi$. The new domain will be denoted by
$\tilde{\Om}$ so that we must solve the non-dimensionalized problem
$$
(i\nabla+\phi A_{N})^{2}\Psi=\mu\Psi\;\hbox{in}\;\tilde{\Om},
$$
where $A_{N}$ satisfies $\nabla\times A_{N}=1\hat{z}$.
When the domain is smooth (i.e. has no corners),
the leading order term in the expansion for $T$ is linear in the flux:
$\frac{T_c-T}{T_c} \sim \frac{\xi^2_0}{R^2}
\mu_0 \Ba $. Here $R=(|\Om|/2\pi)^\half$, and $\mu_0$ is
determined by the eigenvalue problem
\be
\psi_{\eta \eta} +(\mu_0-(\eta-\mu_0^{1/2})^2) \psi=0,
\;\;\psi_{\eta}(0)=0,\; \lim_{\eta \ra \infty}\psi(\eta)=0,\;\psi(0)=1.
\label{degns}
\ee
This equation was introduced by Saint-James and deGennes \cite{stde}
in their work on nucleation of superconductivity near a half plane.

Higher order terms in the expansion are computed through a careful
boundary layer analysis. In general
the wave function concentrates near the
point of the boundary where the curvature $\ka$ is maximal.
It decays exponentially fast away from this point both tangentially
along the boundary of $\Omega$
(on a lengthscale of $O(\phi^{-1/8}R)$) and radially into the
interior of $\Omega$
(on a lengthscale of $O(\phi^{-1/2}R)$) \cite{best}.
The expansion for $\mu$ takes the form
\be
 \frac{T_c-T}{T_c}=
\frac{\xi_0^2}{R^2}(\mu_0 \phi + \mu_1 \phi^{1/2} +\mu_2
\phi^{1/4} +...). \label{expT} \ee Here $\mu_1=-\frac{\ka}{3I_0}$
and $\mu_2=\frac{1}{I_{0}}
\left(\frac{-\sqrt{\mu_{0}}\kappa_{ss}}{6}
\right)^{1/2},$ where $I_0=\int_0^\infty \psi^2$ and $\kappa$ and
$\kappa_{ss}$ are evaluated at the point of maximum curvature (with
$\kappa_{ss}$ assumed to be strictly negative there). Solving
(\ref{degns}) numerically we obtain $\mu_0 = 0.59$ and $I_{0}=1.312$.

In the special case where $\Om$ is a disc, the wave function has
radial symmetry. In fact, it is possible to obtain an explicit
solution to the linearized problem
in terms of Kummer functions \cite{bezw}. However it is
still beneficial to write down the asymptotic expansion because it
captures simple geometric characteristics of the domain. We thus
obtain
\be
\frac{T_c-T}{T_c}= \frac{\xi_0^2}{R^2}(0.59 \phi +\mu_1 \phi^{1/2}
+f(\phi)+...), \label{expD} \ee where $f(\phi)$ is an $O(1)$
oscillatory function of $\phi$ that approaches a periodic function
as $\phi \rightarrow \infty$. We note that a rigorous mathematical
treatment of this problem within the context of nonlinear bifurcation
theory has been carried out in
\cite{bpq}.

When $\Om$ is a square, the curvature function is singular and the
previous expansion is no longer valid. Our computations on the
linearized eigenvalue problem indicate that the wave function now
concentrates near
the corners. A simple analysis shows that the
leading order term in the expansion for $T$ is
\be
\frac{T_c-T}{T_c} \sim \frac{\xi_0^2}{R^2}\mu_c \phi\;\hbox{for}\;
\phi>>1.
\label{expS}
\ee
The coefficient $\mu_c$ is the ground state for the
Schrodinger operator
$(i \nabla + A_{N})^2$ in the quarter-plane.
The eigenvalue can be estimated by an upper bound derived from the Rayleigh
quotient \cite{fdm}. Surprisingly the upper bound computed in \cite{fdm}
is much lower than our numerical computation (see below).

Expansions of the phase transition curve $T(H)$ for low values of
the flux follow from regular perturbation theory.
Let $p$ be the solution of Poisson equation $\nabla^2 p=1$
in $\tilde{\Om}$ with homogeneous Dirichlet boundary conditions, and set
$\zeta_0=-\frac{1}{2\pi} \int_{\tilde{\Om}} p.$
To leading order in $\phi$ we
find
\be
\frac{T_c-T}{T_c}\sim \frac{\xi_0^2}{R^2} \zeta_0 \phi^2.
\label{expL}
\ee
For example, we obtain $\zeta_0=0.25$ for a disc, and $\zeta_0=
0.22$ for a square.

In Figure 1 we present the numerical computations of the values of
$\frac{R^2}{\xi_0^2}\frac{T_c-T}{T_c\phi}$ as a function of $\phi$
for a square (solid line) and for a disc (arrows) with the same
area where $\phi$ lies in the range $[0,16]$. The results are in
perfect agreement with our theory for low $\phi$. It is observed
that superconductivity nucleates for a square at a higher
temperature than for a disc. Our asymptotic results indicate that
the nucleation in a square will be at higher temperature than for
{\em any} smooth domain in the large flux regime.
>From our numerical calculations we estimate $\mu_c =0.55$. As was
mentioned above this is about twice the estimate of \cite{fdm}.
We cannot explain this large discrepancy. We point out, however that
Kato et al. \cite{kem} have solved numercially the time dependent
Ginzburg Landau equation in a square and studied the existence
of superconductivity above $H_{c2}$. They expected to get
the behavior predicted in \cite{stde}. Instead they found what they
called an `anomalous phenomena', which, formulated in our
notation, is equivalent $0.5 < \mu_c < 0.555$. This value is in agreement
with our results.

We compare our expansions with experimental results. We will make a
least-squares fit of the
experimental data to an expansion of the form
\be
\frac{T_c-T}{T_c} \sim \frac{\xi_0^2}{R^2}\mu^0_e \phi
,\,\,\phi \gg 1, \label{experL} \ee and
\be
\frac{T_c-T}{T_c} \sim  \frac{\xi_0^2}{R^2}\zeta_e \phi^2,\;\;
 \phi\ll1. \label{experS}
\ee
 To determine the coefficients $\mu^0_e$ and $\zeta_e$
from a given
experiment we need to know $\xi_0$ and the actual size of the
domain. Therefore
it is interesting to note the {\em ratio} $\dl=\mu^0_e/\zeta_e$ is
independent of both $\xi_0$ and the domain scale. It is controlled
solely by the shape of the domain.

We first consider the experiment of Buisson et al. \cite{bui} who
studied the $T-H$ line for an aluminum disc of radius $7.2\;\mu m
$. For a disc we obtain the theoretical value of the ratio defined
above to be $\dl_0=
.59/\zeta_0=2.36.$ The
experimental value is found to be $\dl_e = 2.26$.

It has been suspected \cite{bui},\cite{bezw} that some of the
anomalies in the experiment were caused by the contact leads that
modified the spherical geometry. Motivated by the theoretical
expansion (\ref{expD}) we shall test this hypothesis by assuming
\be
\frac{T_c-T}{T_c} \sim \frac{\xi_0^2}{R^2}(\mu^0_e \phi+\mu^1_e
\phi^{1/2}),\,\,\phi \gg 1. \label{experL2} \ee
In particular, we wish to
estimate the coefficient of the
$\phi^{1/2}$ term in the expansion (\ref{experL2}) since
this term depends linearly on the maximal curvature.
We used the value $\xi_0 =0.2 \;\mu m$ obtained \cite{bui} from
the low $\phi$ expansion. We found that the radius of curvature at
the nucleation point is $6.5  \; \mu m \pm 0.8 \; \mu m$ which is
in reasonable agreement with the actual radius of $7.2\;\mu m$. Thus we
conclude that at least at the high $\phi$ regime the results
were not affected by the contact leads.

Moshchalkov et al. \cite{mos} performed a similar experiment with
a $1 \times 1\; \mu m$ aluminum square. They used fluxes of up to 4
flux quanta. The experimental data was fitted to the
numerical solution of the associated eigenvalue problem for a disc
with area 1. For this purpose they chose $\xi_0
=0.1\;\mu m$. A similar comparison was later performed in
\cite{bezw}. A recent experiment from the same laboratory \cite{bru2}
provides more extensive data. In this experiment a $2 \times 2\;
\mu m$ square was used. The authors measured the phase transition
curve for fluxes of up to 16 flux quanta and again compared the data
to the expected theoretical values for a disc of comparable area.

We shall compare the data of \cite{bru2} for the square to the
asymptotic theoretical formulas (\ref{expS}) and (\ref{expL}).
Again this comparison is complicated by the need to estimate the
value of $\xi_{0}$. As a first attempt, we estimate the value of
$\xi_0$ by comparing the experimental data for low values of
$\phi$ to the theoretical expansion (\ref{expL}) and obtain $\xi_0
= 0.095 \;\mu m$. However, using this value, the experimental
results \cite{bru2} predict $\mu_e^0 > 0.81$ in the large $\phi$
expansion (\ref{experL}). This is way above even the theoretical
value for the disc. In fact, there is no $\xi_{0}$ that allows one
to accurately match the theoretical curve to the experimental
data. To demonstrate this point, we note that our theory gives the
estimate $\dl = 2.5$, while the experimental estimate is
$\dl=3.7$. (Our estimate of $\dl$ is based upon our numerical
evidence that $\mu_{c}$ from expansion (\ref{expS}) is
approximately 0.55) We point out that other laboratories have
recently reported \cite{zhpr}, \cite{dav}, \cite{bui} values for
$\xi_0$ for aluminum near $0.2 \;\mu m$.

One plausible explanation for the deviation lies in the importance
of the nonlocal effects due to the electric contacts \cite{str}.
In the low $\phi$ regime the coherence length is large compared to
the size of the square, so that wave function, and thus the
transition temperature, are affected by the neighborhood of the
square. Indeed Figure 3 of \cite{str} indicates that the electric
contacts do affect the phase transition curve for low $\phi$, and
their effect diminishes as $\phi$ increases. Therefore, we shall
estimate $\xi_{0}$ using the high $\phi$ data. We thus obtain
$\xi_0 = 0.1195  \;\mu m$. The circles in Figure 1 corresponds to
the experimental data of \cite{bru2} with this choice of
$\xi_{0}$. The nonlocal effect was apparently less pronounced in
\cite{bui} since their domain was 10 times larger.

An examination of the three curves in Figure 1 reveals that while
the experimental curve lies well below the theoretical curves, the
{\em location of the peaks} in the experimental curve perfectly match the
location of the
peaks of the theoretical curve for the square. On the other
hand, the experimental peaks deviate considerably from the
theoretical peaks for the {\em disc}. Bruyndoncx et al.
\cite{bru2} introduced the notion of `effective square area'
(which is smaller than the true geometrical area) in order to
obtain a fit between the experimental data and the theory for the
disc. We have shown here that the actual experimental peaks are in
perfect agreement with the theory for the square. We also point
out that while the experimental slope for low $\phi$ (near 0.5) is
about 0.13 (well below the theoretical 0.22 value), the
experimental curve steepens just before the first peak ($\phi$
near 1.7), and the local slope is about 0.19. It still remains to
explain the deviation between the experimental and theoretical
curves for intermediate values of $\phi$, both in actual values
and in the amplitude of the oscillations. A possible explanation
is in the way in which the transition temperature is defined.
Bruyndoncx et al. \cite{bru2} set it to be the temperature for
which the resistivity drops to half the normal value. As the
applied flux increases, the temperature drops, and the uncertainty
interval shortens.

To summarize, we have developed general formula (\ref{expT}),
(\ref{expD}) and (\ref{expS}) for the asymptotic transition line
in mesoscopic superconducting samples in the limit of large flux.
We further derived a formula (\ref{expL}) for the asymptotics of
that curve in the limit of low flux. In the large flux limit and
for smooth domains, our formula captures the universal linear base
curve that relates to the fact that the wave function concentrates
near the boundary which appears locally as a half plane. It also
contains terms proportional to fractional powers of the applied
flux that depend on universal parameters and on the local geometry
of the sample's surface. In the special case of a disc our formula
resolves an oscillatory dependence of $T$ upon $\phi$. We have
also shown that a square sample exhibits a different behavior from
smooth domains already at the linear leading order term. Finally
we used the new theory to analyze recent experiments.

\paragraph{Acknowledgement}
We are grateful to V. Bruyndoncx and O. Buisson for sharing with
us a wealth of experimental data, and to J. Berger for bringing refs.
\cite{fdm} and \cite{kem} to our attention.

\medskip\noindent
H.T.J: Department of Mathematics, Indiana University, Bloomington,
IN 47405, USA. E-mail: hjadalla@indiana.edu

\medskip\noindent
J.R.: Department of Mathematics, Technion, Haifa 32000, Israel.
E-mail: koby@math.technion.ac.il. Research supported in part by
the U.S.-Israel Binational Science Foundation, by the Israel
Science Foundation and by the Posnansky Research Fund in high
temperature superconductivity.

\medskip\noindent
P.S.: Department of Mathematics, Indiana University, Bloomington,
IN 47405, USA. E-mail: sternber@indiana.edu. Research supported in
part by the N.S.F. under grant DMS-9322617 and a U.S.-Israel
Binational Science Foundation Grant.



\begin{thebibliography}{99}

\bibitem{bpq} P. Bauman, D. Phillips and Q. Tang, Arch. Rat. Mech. Anal.
{\bf 142}, 1, 1998.

\bibitem{bezw} R. Benoist and W. Zwerger Z. Phys. {\bf B103}, 377 (1997).

\bibitem{beru} J. Berger and J. Rubinstein, Phys. Rev. Lett. {\bf 75},
320 (1995).

\bibitem{best} A. Bernoff and P. Sternberg, J. Math. Phys. {\bf 39},
1272, (1998).

\bibitem{bru2} V. Bruyndoncx, J.G. Rodrigo, T. Puig, L. Van Look,
V.V. Moshchalkov and R. Jonckheere, preprint.

\bibitem{bui}  O. Buisson, P. Gandit, R. Rammal, Y.Y. Wang and
B. Pannetier, Phys. Lett. {\bf 150}, 36 (1990).

\bibitem{dav}  D. Davidovich et al. Phys. Rev. B {\bf 55}, 6518 (1997).

\bibitem{fdm} V.M. Fomin, J.T. Devreese and V.V. Moshchalkov,
 Europhys. Lett. {\bf 42}, 553 (1998).

\bibitem{deg} P.G. de Gennes {\em Superconductivity in metals and Alloys},
Addison Wesley (1989).

\bibitem{grpa} R.P. Groff and R.D. Parks, Phys. Rev. {\bf 176}, 567 (1968).

\bibitem{kem} R. Kato, Y. Enomoto and S. Maekawa, Phys. Rev. B {\bf 47},
8016 (1993).

\bibitem{lipa}  W.A. Little and R.D. Parks,
Phys. Rev. Lett {\bf 9}, 9 (1962).

\bibitem{mos} V.V. Moshchalkov, L. Gielen, C. Strunk,
R. Jonckheere, X. Qiu, C. Van Haesendonck and Y. Bruynserade,
Nature (London) {\bf 373}, 319 (1995).

\bibitem{stde} D. Saint-James and P.G. deGennes, Phys. Lett. {\bf
7}, 306 (1963).

\bibitem{str} C. Strunk, V. Bruyndoncx, V.V. Moshchalkov, C. Van
Haesendock, Y. Bruynserase and R. Jonckheere, Phys. Rev. {\bf B 54},
R12701, (1996).

\bibitem{zhpr} X. Zhang and J.C. Price, Phys. Rev. B {\bf 55}, 3128 (1997)

\end{thebibliography}
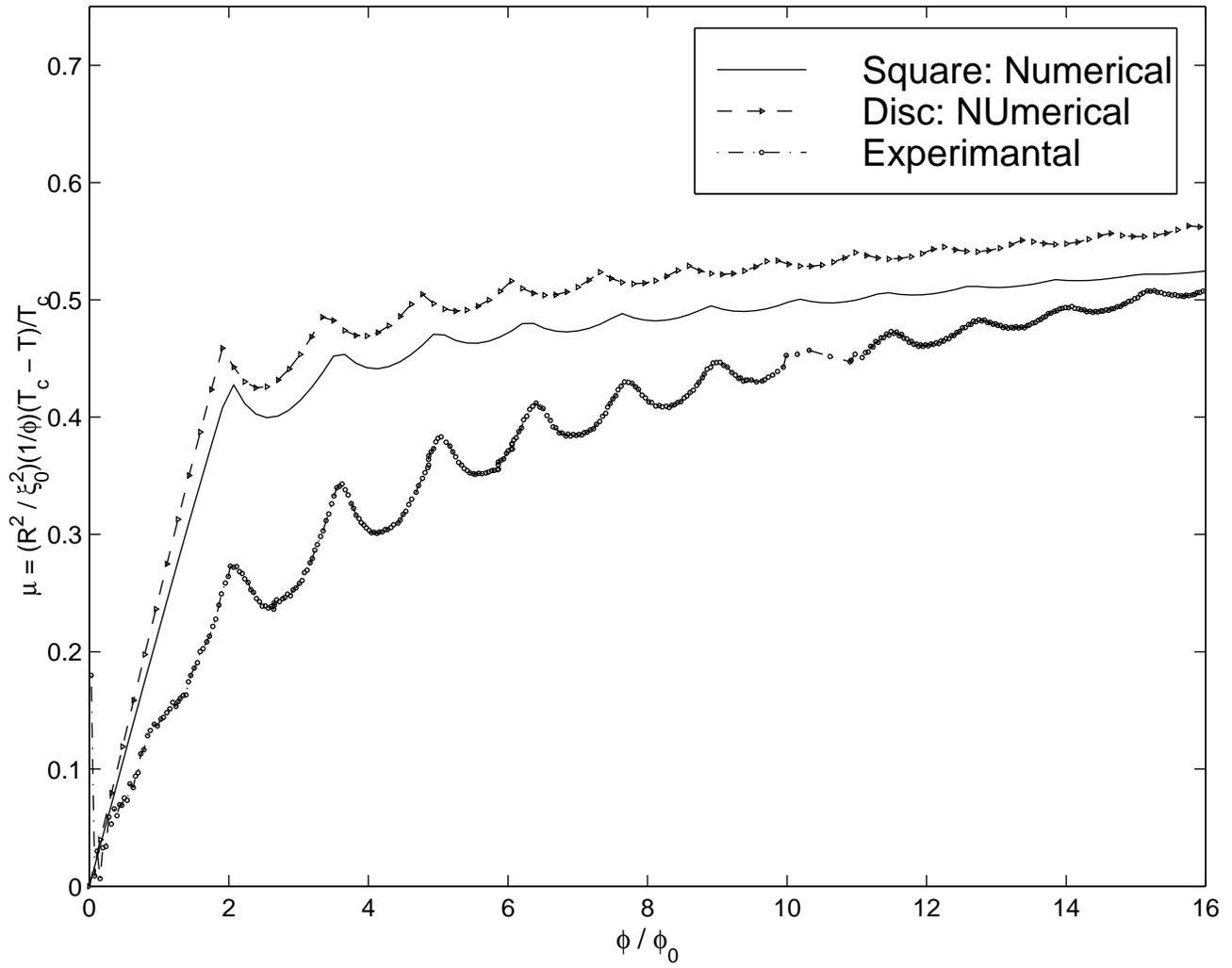
\end{document}